\documentclass[a4paper,11pt]{article}
\pdfoutput=1
\usepackage{jheppub}

\usepackage[utf8]{inputenc}
\usepackage{comment}
\usepackage{graphicx,rotating,amsmath,amsfonts,amssymb}
\usepackage[svgnames]{xcolor}
\usepackage{subfigure}
\usepackage{enumerate}
\usepackage{mathrsfs}

\usepackage{soul}
\setstcolor{red}

\usepackage{tikzexternal}
\usepackage{tikz}
\usetikzlibrary{external}

\newcommand{\be}{\begin{equation}}
\newcommand{\ee}{\end{equation}}

\newcommand{\mpl}{M_{\rm pl}}

\newcommand{\td}{\mathrm{d}}

\DeclareMathOperator{\atan}{arctan}
\DeclareMathOperator{\acos}{arccos}

\title{Wormholes in conformal gravity}

\author[a]{M. Hohmann,}
\author[a]{C. Pfeifer,}
\author[b]{M. Raidal,}
\author[b,c]{and H. Veerm\"ae}

\affiliation[a]{Institute of Physics, University of Tartu, W. Ostwaldi Str 1, 50411, Tartu, Estonia}
\affiliation[b]{National Institute of Chemical Physics and Biophysics,  R\"avala 10, 10143 Tallinn, Estonia}
\affiliation[c]{Theoretical Physics Department, CERN, CH-1211 Geneva 23, Switzerland}

\emailAdd{martti.raidal@cern.ch}
\emailAdd{manuel.hohmann@ut.ee}
\emailAdd{christian.pfeifer@ut.ee}
\emailAdd{hardi.veermae@cern.ch}

\abstract{We present a new class of solutions for static spherically symmetric wormhole spacetimes in conformal gravity and outline a detailed method for their construction.  As an explicit example, we construct a class of traversable and non-traversable wormholes that are  locally conformal to Schwarzschild-(anti) de Sitter spacetimes. These wormhole spacetimes are exact vacuum solutions in, but not being limited to, Weyl gravity and conformal scalar-tensor theories. Importantly, the method implies that every conformal theory of gravity with static spherically symmetric solutions will trivially contain wormholes without the need for exotic matter. Applying those results on gravitational theories that possess conformal symmetry in the ultraviolet regime, the central singularities of black holes can be replaced with wormhole throats. We speculate on possible phenomenological consequences. We also discuss the inclusion of matter fields and give explicit examples of charged wormholes in Weyl gravity and conformal scalar-tensor gravity.
}

\begin{document}
\begin{flushleft}
	\hfill		  CERN-TH-2018-026
\end{flushleft}
\maketitle
\flushbottom

\section{Introduction}
\label{sec:in}

Although Nature is not conformally invariant, there are good theoretical reasons to believe that  conformal symmetry might play an important role in ultraviolet completions of gravity (for the discussion see, e.g.,~\cite{Maldacena:2011mk}). Explicit attempts in this direction have been proposed~\cite{Salvio:2014soa,Modesto:2016max} that may represent prototypical theories of gravity valid up to infinite energy. The conformal invariance must be broken at least by quantum effects through anomalies~\cite{Capper:1974ic}, yet the theory may be an ultraviolet fixed point of scale invariant gravity~\cite{Salvio:2017qkx}. As a result, some features of the conformal gravity may survive at low energies as manifestations of the unbroken high-energy phase of the theory.

A fascinating phenomenon in the geometric description of gravity is the potential existence of wormholes \cite{Visser:1995cc}. The first wormhole solution known was the Einstein-Rosen bridge \cite{Flamm1916,PhysRev.48.73} in the maximally extended Schwarzschild solution. It encouraged speculations about the possibility of wormhole spacetimes which allow to connect distant regions by a shortcut. However, it turned out that the Einstein-Rosen bridge is non-traversable and that this property is shared by all wormholes satisfying the Einstein equations, because a traversable wormhole necessarily violates the null energy condition, which is satisfied for all known types of matter~\cite{Morris:1988cz, Morris:1988tu, PhysRevLett.81.746}.

In modified theories of gravity the situation can be different~\cite{Harko:2013yb,Lobo:2008zu,Capozziello:2012hr,Bambi:2015zch,Varieschi:2015wwa,Duplessis:2015xva,Nascimento:2017def,Sahoo:2017ual,Zubair:2017oir,Kuhfittig:2016mwd,Kuhfittig:2018vdg,Barcelo:1999hq,Willenborg:2018zsv,Shaikh:2015oha,Shaikh:2016dpl,Shaikh:2017zfl} and wormholes may be constructed {\it without} exotic matter. Wormholes in the vacuum of quadratic gravity have been found by noting that any metric with vanishing Ricci scalar $R$ is a vacuum solution of the theory and then solving the equation $R=0$~\cite{Duplessis:2015xva,Shaikh:2016dpl}. Traversable wormholes with no matter sources have also been found in Chern-Simons modified gravity~\cite{Nascimento:2017def}. Wormhole spacetimes in the context of violation of the null energy condition by the surrounding matter have been studied in conformal Weyl gravity~\cite{Lobo:2008zu,Varieschi:2015wwa}, and a static spherically symmetric wormhole in the vacuum of Weyl gravity was found in Ref.~\cite{Oliva:2009zz}. Wormhole solutions have also been studied in conformally coupled scalar-tensor theories \cite{Barcelo:1999hq,Prester:2014voa,Willenborg:2018zsv}.

In this article we present a general prescription of how to construct a new class of traversable as well as non-traversable wormholes from spherically symmetric spacetimes by a cut and conformal transformation procedure. To be more precise, we take a hypertube part of the spacetime in consideration and apply a \emph{local} conformal transformation to it. We derive the conformal factor which turns this part of spacetime into a wormhole spacetime. This method can be used to construct new solutions in \emph{any} conformal theory of gravity. Notably, because of the cutting procedure, these spacetimes are \emph{globally} inequivalent even in the context of conformal theories.

Our work reveals a new, so far largely overlooked, feature of conformal gravity theories, i.e., theories whose field equations are invariant under the conformal rescaling
of the metric,
\be
	g_{\mu\nu} \to \Omega(x)^{2} g_{\mu\nu} .
\ee
We demonstrate that theories containing a spherically symmetric solution, and this can be expected for all conformal theories of gravity, also have solutions which are traversable wormholes. The key point is that in those theories improper conformal transformations, with $\Omega(x)$ zero or infinite at some points and thus not well defined globally, can be formally used to construct new solutions. As a result, the existence of wormhole vacua in conformal theories of gravity is a completely general, unavoidable feature of conformal gravity. We stress, however, that our method is also applicable to non-vacuum solutions of conformal theories containing matter.

Gravitational theories with conformal symmetry make frequent appearances in the literature of gravity. Perhaps the most well known example is Weyl gravity~\cite{Weyl:1917gp,Weyl1918,Weyl:1918ib,Weyl:1919fi} whose action is the square of the Weyl tensor. It is an example of quadratic gravity which has the advantage that it is, in contrast to general relativity, renormalizable~\cite{Stelle:1976gc}. However, as a higher derivative theory it suffers from the Ostrogradsky instability in the form of ghosts~\cite{Ostrogradsky:1850fid,Woodard:2015zca}. Attempts to avoid the dangerous ghosts are numerous~\cite{Pauli:1943,Dirac:1942,Smilga:2005gb,Bender:2007wu,Grinstein:2007mp,Bender:2008gh,Maldacena:2011mk,Chen:2012au,Salvio:2015gsi,Raidal:2016wop,Narain:2016sgk,Narain:2017tvp,Donoghue:2017fvm} but so far no widely accepted solution exists. Another well known example is scalar-tensor theory with the scalar field conformally coupled to the Ricci scalar. The latter is sometimes referred to as a fake conformal theory since the conformal symmetry acts as a gauge symmetry by which we can remove the scalar degree of freedom. As a result the theory is equivalent to general relativity, at least locally. This example demonstrates the generality of our proposal and its applicability to theories without pathologies such as ghosts.

Some solutions obtainable by our procedure have already been presented in literature. For example, the wormholes in~\cite{Oliva:2009zz} represent a class of wormhole spacetimes proposed in this paper. Interestingly, the non-conventional black holes in Weyl gravity~\cite{Mannheim:1988dj} are obtained from our construction in the limit of vanishing wormhole throat radius.  We stress that neither~\cite{Oliva:2009zz} nor~\cite{Mannheim:1988dj} used conformal symmetry to obtain their solutions, although it has been noted that these solutions are locally conformal to the Schwarzschild-(anti) de Sitter (S-(A)dS) black holes. This is, in fact, true for all static spherically symmetric vacuum solutions of both Weyl and conformal scalar-tensor gravity.

Although the main aim of this work is to demonstrate the existence of wormhole solutions in conformal gravity, it is tempting to speculate on possible theoretical and phenomenological consequences of our proposal. If conformal symmetry is restored in the ultraviolet regime of gravity, while the low energy regime is described by general relativity~\cite{Salvio:2014soa,Salvio:2017qkx}, our results may have implications for black hole physics. As a result, in the vicinity of the centre of a black hole, corresponding to the ultraviolet regime, the classical singularity can be removed and replaced by a wormhole throat through quantum effects. A similar idea has been discussed in \cite{Bambi:2016wdn}.

Alternatively, the wormholes which appear in conformal theories may survive the conformal symmetry breaking and exist at low energies. Their properties mimic the ones of  black holes if only long-distance gravitational effects are considered. However, the absence of a horizon allows them to be distinguished from black holes~\cite{Cardoso:2016rao}, for example, in the coalescence of two of those objects, when a fraction of their total mass is radiated by gravitational waves, as observed by the LIGO experiment~\cite{Abbott:2016blz,Abbott:2016nmj}. If these coalescing objects are wormholes instead of black holes, gravitational wave echoes~\cite{Cardoso:2017cqb} following the ringdown phase could be present. The absence of a horizon also changes drastically the radiation from those objects~\cite{Damour:2007ap, Berthiere:2017tms}. We note, that modifications to Hawking radiation due to conformal invariance can be expected even in the presence of an horizon~\cite{Bambi:2016yne}. If large primordial curvature fluctuations would collapse~\cite{Hawking:1971ei} into primordial wormholes instead of black holes, those light primordial wormholes do not evaporate during the Hubble time and could thus exist today. This implies that the astrophysical constraints on those objects, summarised in~\cite{Carr:2017jsz}, should be revised and the unknown dark matter of the Universe might consist of light primordial wormholes. More speculatively but also more fundamentally, wormholes may play a role in the understanding of quantum entanglement~\cite{Maldacena:2001kr,Maldacena:2013xja}.

This article is outlined as follows: In Section~\ref{sec:solution} we briefly review conformal transformations acting between static spherically symmetric spacetimes and present the construction of wormhole spacetimes. Vacuum wormhole solutions of Weyl gravity and conformal scalar-tensor gravity will be discussed in Section~\ref{sec:ConfGrav}, along with some concrete examples. We briefly discuss non-vacuum solutions and, to provide specific examples, we derive charged wormhole solutions in Weyl gravity and conformal scalar-tensor gravity minimally coupled to the electromagnetic field. We conclude in Section~\ref{sec:end}. Some technical details related to our solutions are collected in Appendix~\ref{A}. Throughout the article we use natural units $\hbar = c = 1$ and we use the sign conventions $(+,+,+)$ as defined in \cite{Misner:1974qy}.

\section{Construction of the wormhole metric}
\label{sec:solution}

We construct the wormhole from black hole spacetimes using local conformal transformations. The method restricts spacetime to a hypertube and then conformally extends that hypertube to form a wormhole spacetime. The conformal factor is chosen such that it becomes singular at the edges of the hypertube and thus there is a connection between the restriction and the conformal factor. The method may also be applicable to non-spherical spacetimes, but to demonstrate the mechanism we restrict our treatment to the spherical ones.

In the following we will first recall that in conformal theories of gravity the most general static spherically symmetric metric is of the Kerr-Schild type, after which we will demonstrate how such metrics can be transformed to wormholes.

\subsection{Spherically symmetric metrics and conformal transformations}

Given a suitable choice of coordinates static, spherically symmetric spacetimes can always be cast in the form
\be\label{stat:g_sss}
g =	- a(r)\, \td t^2 + b(r)^{-1}\td r^2 + r^2 \td \mathbf{\Sigma}^2,
\ee
where $a(r)$ and $b(r)$ are functions of the radial coordinate only and $\td \mathbf{\Sigma}$ is the canonical length element on a $D-2$ dimensional sphere. The construction of the wormhole only assumes a spacetime dimension $D\geq3$ in order for the wormholes to exist. In conformal gravitational theories the metric \eqref{stat:g_sss} can be reduced further and, as we will show, it is sufficient to find solutions satisfying an additional ansatz, e.g., $a = b$, which we will refer to as a Kerr-Schild type static spherically symmetric metric.\footnote{A static spherically symmetric metric can be expressed in the Kerr-Schild form $g_{\mu\nu} = \eta_{\mu\nu} + \phi k_{\mu}k_{\nu}$, with $\eta^{\mu\nu}k_{\mu}k_{\nu} = 0$. This form exists precisely if $a = b$ \cite{PhysRevD.94.044023,Stephani}, thus the name.}

A conformally transformed metric $\tilde g$ with spherically symmetry is obtained from the original metric $g$ by
\be
	\tilde g = \Omega(r)^2 g,
\ee
where the conformal factor $\Omega(r)$ is in principle an arbitrary function of $r$. In order to bring it to standard spherically symmetric form \eqref{stat:g_sss} again one must introduce a new coordinate $\tilde r = \Omega r$ and finds
\be\label{eq:spericalconf}
	\tilde g = - \tilde a(\tilde r)\, \td t^2 + \tilde b(\tilde r)^{-1}\td\tilde  r^2 + \tilde r^2 \td \mathbf{\Sigma}^2,
\ee
where the new functions $\tilde{a}(\tilde{r})$ and $\tilde{b}(\tilde{r})$ read
\be\label{stat:g_sss_c}
	\tilde a(\tilde r) =  \Omega^{2}a(\tilde r/\Omega), \qquad
	\tilde b(\tilde r) =  \frac{\Omega^{2}b(\tilde r/\Omega)}{(\Omega - \tilde  r\frac{\td\Omega}{\td\tilde r})^{2} }\,,
\ee
and $\Omega$ is now treated as a function of $\tilde r$. This is the conformal transformation rule of static spherically symmetric metrics.

It is always possible to find a conformal transformation such that $\tilde a = \tilde b$ by solving the equations
\be\label{eq:toKS}
	\pm \sqrt{\frac{a(r)}{b(r)}} = \Omega^{-2} \frac{\td \tilde r }{\td r}  = - r^2 \frac{\td (\Omega r)^{-1} }{\td r}
\quad \Leftrightarrow  \quad
	\Omega(r) = \pm \left(r  \int^{r}_{r_0} \sqrt{\frac{a(R)}{b(R)}} \frac{\td R}{R^2} \right)^{-1}.
\ee
 Thus for conformally invariant theories of gravity it suffices to consider a Kerr-Schild type metric
\be\label{eq:kerr-schild}
	\tilde g = - \tilde a(\tilde r) \td t^2 + \tilde a(\tilde r)^{-1} \td\tilde r^2 + \tilde r^2 \td\mathbf{\Sigma}^2,
\ee
to find the spherically symmetric solutions of the theory. All other static spherically symmetric solutions can be obtained form it by means of conformal transformations.

Note, however, that the choice $\tilde a = \tilde b$  does not completely remove the conformal degeneracy. Taking \eqref{stat:g_sss_c} and imposing $a = b$ and $\tilde a = \tilde b$ yields the condition for conformal transformations between Kerr-Schild type metrics,
\be\label{eq:omega_d0}
	\left( \Omega - \tilde  r\frac{\td\Omega}{\td\tilde r} \right)^{2} = 1
	\qquad \Leftrightarrow  \qquad
	\Omega = \frac{\tilde  r}{r_{*}} \pm 1 = \frac{\pm1}{1- r/ r_{*} },
\ee
where $r_{*}$ is an arbitrary constant. Thus the choice $\tilde a = \tilde b$ restricts the infinite dimensional local conformal symmetry to a group of one parameter transformations.

We will omit the tilde for the rest of the paper, because we can assume $a=b$ without loss of generality.

\subsection{The wormhole transformation}

Having found solutions of a conformal theory of gravity of the form \eqref{eq:kerr-schild} other solutions can be constructed via conformal transformation and analytic continuation/restriction of the solutions. To eventually obtain a wormhole metric, we apply again a conformal transformation to the metric \eqref{eq:kerr-schild} with the aim to map it to the form
\be\label{stat:g_sss_wh}
	\hat g =  \Omega^2  g = - \hat a(l) \, \td t^2 + \hat a(l)^{-1} \td l^2 + (l^2 + d^2) \td \mathbf{\Sigma}^2.
\ee
For $\hat{a}(l)$ being positive and continuous around $l = 0$ this spacetime describes a stationary wormhole with a throat radius $d$.\footnote{Allowing negative values of $d$, the throat radius is given by $|d|$.} The properties of the throat of the wormhole \eqref{stat:g_sss_wh} are briefly outlined in Appendix~\ref{ssec:throat} in the Morris-Thorne formalism.

Repeating the steps from the previous section and using  Eq.~\eqref{stat:g_sss_c}, a conformal transformation of a Kerr-Schild type metric $ g$ with a conformal factor $ \Omega$ yields
\be\label{eq:ghat}
	\hat g = - \hat a(\hat r)\, \td t^2 + \hat b(\hat r)^{-1}\td\hat  r^2 + \hat r^2 \td \mathbf{\Sigma}^2,
\ee
with $\hat r =  \Omega  r$ and
\be\label{eq:hatcoef}
	\hat a(\hat r) =   \Omega^{2} a(\hat r/ \Omega), \qquad
	\hat b(\hat r) =  \frac{ \Omega^{2}  a(\hat r/ \Omega)}{\big( \Omega - \hat r \frac{\td  \Omega}{\td\hat r}\big)^{2}} \, .
\ee
The main difference to the previous subsection is that we have assumed $ a =  b$ in the metric~\eqref{eq:kerr-schild} from which we started. Introducing a new coordinate $l$ via $\hat r^2 = l^2 + d^2$ in~\eqref{eq:ghat} for some fixed constant $d$ and demanding that the resulting metric takes the form~\eqref{stat:g_sss_wh} yields the following condition on the conformal factor $ \Omega$
\be
	\frac{\hat a}{\hat b} = \bigg( \Omega - \hat r \frac{\td \Omega}{\td\hat r}\bigg)^{2} = \bigg(\frac{\td l}{\td \hat r}\bigg)^2,
\ee
which can be solved explicitly for $ \Omega$\footnote{The general solution is determined up to a sign, that is  $\eta  \equiv \eta_0 \pm\atan\left(l/d\right)$. However, as the sign in $\eta$ can be flipped by $l \to -l$, then we pick the solution with the negative sign without loss of generality.} as
\be\label{eq:omega_wh}
	 \Omega
	= \eta(l)\sqrt{1+\frac{l^2}{d^2}}
	= \frac{d}{ r \cos\left(\eta_0 - d/ r\right)},
\ee
where we defined $\eta(l) \equiv \eta_0 - \atan\left(l/d\right)$ and  $\eta_0$ is an integration constant corresponding to the conformal factor at the throat of the wormhole, $l=0$. The coordinate transformation is given by
\be\label{r(l)}
	r(l) = \frac{d}{\eta(l)} \qquad \Leftrightarrow \qquad
	l(r) = d \tan\left(\eta_0 - d/r\right) .
\ee
This implies that for any Kerr-Schild solution \eqref{eq:kerr-schild} there exists a class of wormhole solutions \eqref{stat:g_sss_wh} with
\be\label{hat_a}
	\hat a(l) = \left(1 + \frac{l^2}{d^2}\right)\, \eta(l)^2\  a(d\, \eta(l)^{-1}).
\ee
These solutions are parametrised by the throat radius $d$ and the parameter $\eta_0$ contained in $\eta(l)$.

Observe that the range of the coordinates $l \in (-\infty,\infty)$ does \emph{not} cover the region $ r \in  (0, \infty)$ for the initial spacetime metric \eqref{eq:kerr-schild} and in some cases it is necessary to extend the metric so that the radial coordinate takes values on the projectively extended real numbers. We stress that this is not an artefact of the coordinate transformation as, by construction, the conformal transformation \eqref{eq:omega_wh} diverges at the boundary $ r(l \to \pm \infty)$.  The domain of the transformation, i.e., the region of the initial space where the conformal factor is well defined, is implicit in the coordinate transformation \eqref{r(l)} and will be discussed in detail in the next subsection.

\subsection{Domain of the transformation}

As stated before, to understand on which regions of the original spacetime the conformal factor is well defined and thus which regions are mapped into the wormhole spacetime, it is sufficient to study the coordinate transformation $r(l) = d/(\eta_0 - \atan\big(l/d\big))$. The domain of the transformation is a hypertube, i.e., a restriction of the radial coordinate to an interval and no restrictions on the time or the angular coordinates, so the spatial slice at constant $t$ is simply a spherical shell of finite thickness. In 2+1 dimensions one may visualise this spacetime as a cylindrical shell, a tube. It is described by
\be\label{r_pm}
	 r_{\pm} \equiv  r(l  \to \pm \infty) =  \frac{d}{\eta_0 \mp \pi/2},  \qquad  r_{0} \equiv   r(l=0) = d/\eta_0,
\ee
where $r_{\pm}$ represent the boundaries of the spatial spherical shell. They are mapped to the spatial infinity of the wormhole spacetime. The sphere at $r = r_{0}$ is mapped to the throat of the wormhole. The relation
\be\label{r_pm_d_eta0}
	d = \pi \frac{  r_{+}  r_{-}}{ r_{+} -  r_{-}}, \qquad \eta_0 = \frac{\pi}{2}\frac{ r_{+} +  r_{-}}{ r_{+} -  r_{-}},
\ee
implies, that the transformation \eqref{eq:omega_wh} can be equivalently characterised in terms of $ r_{\pm}$. According to \eqref{r(l)} $r$ can become infinite or even negative for $|\eta_0| < \pi/2$. This is not an issue, if $a(r)$ can be extended from $r \in (0,\infty)$, to the projectively extended real line $\mathbb{R}^{*} = \mathbb{R} \cup \{\infty\}$. This is the one point compactification of the real line, where the positive and negative infinity are joined by adding a single point, $\infty$ (and identifying $\infty = -\infty$)~\cite{Alexandroff1924}. In the rest of the paper $r \in \mathbb{R}^{*}$ will be assumed when necessary.

In a $D$ dimensional spacetime the extension $r \in \mathbb{R}^{*}$ can be understood as gluing of the spheres at $r = \infty$, and $r= -\infty$ at each fixed time $t$. Our goal is to find another solution of the field equation, so the resulting $\hat a$ should be finite, continuous and differentiable at the point the $r = \infty$ is mapped into. This implies that $a(r)$ must have identical asymptotic behaviour for $r \to \pm \infty$.

Consider the case when the point at spatial infinity of the initial spacetime is mapped to a point in the wormhole. If this point is not surrounded by a horizon, that is $\hat a(l(r\to\pm \infty))>0$, then it follows from \eqref{hat_a} that $a(r) \propto r^2$ when $r \to \pm \infty$.  In other words, when the initial spacetime is asymptotically AdS, then it is possible to travel through the point $l(r \to \pm \infty)$, and thus connect the positive and negative spatial infinity of the initial spacetime with a timelike path. Otherwise, if $a(r) \propto r^\alpha$ with $\alpha < 2$, then the spatial infinity of the initial space is turned into a horizon of the wormhole. If the original spacetime is asymptotically dS, then $\hat a(l(r \to \infty)) < 0$, so this point will be surrounded by horizons (assuming that $\hat a < 0$ does not hold everywhere.)

Depending on the value of $\eta_0$ we can identify the following special cases.
\begin{itemize}
\item If $\eta_0 \geq \pi/2$ or equivalently $0 <  r_{-} <  r_{+}$, then the wormhole is obtained from the hypertube determined by $ r \in ( r_{-},  r_{+})$. When the inequality is saturated, $\eta_0=\pi/2$, then $r_{+} = \infty$.

\item  If $\eta_0 \leq -\pi/2$ or equivalently $ r_{-} <  r_{+} < 0$, then the initial spacetime needs to be extended to negative radial coordinates. In this case the wormhole is obtained from the hypertube of the extended solution given by $ r \in ( r_{-},  r_{+})$. When the inequality is saturated, that is $\eta_0=\pi/2$, then $r_{-} = -\infty$.

\item If $|\eta_0| < \pi/2$ or equivalently $ r_{+} < 0 <  r_{-}$, the wormhole is obtained from the region by $ r \in \mathbb{R}^{*}\setminus [r_{+},r_{-}]$ and continued through the spatial infinity.  The spatial infinity of the initial spacetime corresponds to the point where $\eta = 0$. A finite $\hat a$ at that point thus corresponds to $a( r \to \pm\infty) \propto  r^{2}$. Such initial spacetimes are asymptotically (A)dS.

\end{itemize}
A compact way of denoting the range of $r$ is making use of the fact that $r$ lives on $\mathbb{R}^{*}$ and when $r_{+} < r_{-}$ define the interval $(r_{-},r_{+})$ as $\mathbb{R}^{*}\setminus[r_{+},r_{-}]$ as it is commonly done. Thus, on $\mathbb{R}^{*}$, all special cases listed above can be summed up by
\be
	r \in (r_{-},r_{+})
\ee
for arbitrary $r_{\pm} \in \mathbb{R}^{*}$. It also contains the case $r_{-} < 0 < r_{+}$ not listed above. Although $r_{-} < 0 < r_{+}$ corresponds to  a negative throat radius by \eqref{r_pm_d_eta0}, it can result in a well defined wormhole metric and should thus not be excluded.

Another class of special cases are obtained in the limit $r_{\pm} \to 0$ which implies a vanishing throat radius by \eqref{r_pm_d_eta0}. Thus no wormhole is formed. Before looking at the limit $d \to 0$, however, consider the case $d=0$. As the metric ansatz \eqref{stat:g_sss_wh} now reduces to the Kerr-Schild ansatz \eqref{eq:kerr-schild} this transformation maps Kerr-Schild type spacetimes into Kerr-Schild type spacetimes and the transformation is thus given by \eqref{eq:omega_d0} which we repeat here for convenience,
\be
	 \Omega_{d = 0} = \frac{l}{ r_{*}} \pm 1 = \frac{\pm 1}{1 -  r/ r_{*}}.
\ee
Again, $r_{*}$ is an arbitrary constant. Since this conformal transformation has a pole at $r =  r_{*}$, the sphere at $r =  r_{*}$ is mapped to spatial infinity, so the initial spacetime has to be defined on $ r \in \mathbb{R}^{*}$. This conformal factor is also found in the limit $d \to 0$. In that case $ r_{*}$ is either $ r_{+}$ or $ r_{-}$.  For example, assume $l > 0$. The asymptotics of the conformal factor~\eqref{eq:omega_wh} in the limit $d \rightarrow 0_{+}$ then give $ \Omega = l/ r_{+} - 1$, so $ r_{*} =  r_{+}$. The limit $d \rightarrow 0_{+}$ has to be taken by keeping $r_{+}$ constant which implies that $r_{-} \to 0$.

We remark that the consecutive application of transformations $ \Omega_{d = 0}$ and $ \Omega$ is equivalent to a shift $\eta_0 \to \eta_0 \pm d/ r_{*}$.

We have demonstrated that conformal transformations can turn hypertubes cut from general spherically symmetric spacetimes \eqref{stat:g_sss} into traversable wormholes \eqref{stat:g_sss_wh}. The only requirement is the mild condition that $\hat a(l) $ is continuous and positive around the throat, at $l=0$. The wormhole throat in the more conventional Morris-Thorne metric is discussed in Appendix~\ref{ssec:throat}. Traversability of the wormhole is guaranteed if there are no horizons in the region of the initial spacetime which gets conformally transformed to the wormhole spacetime.

Hence \emph{whenever} a conformal theory of gravity has a spherically symmetric solution, it also has wormhole solution that can be constructed from it. We now turn to a specific class of examples in well known gravitational theories with conformal symmetry.

\section{Wormholes in Weyl and conformal scalar-tensor gravity}
\label{sec:ConfGrav}

The following class of wormhole solutions exist in two concrete models of conformal gravity: Weyl gravity and scalar-tensor gravity with a conformally coupled scalar. As the wormholes result form the conformal nature of the theory, they exist even in the absence of matter. Both theories have the S-(A)dS black hole as a solution, for which the metric can be given by \eqref{stat:g_sss} with
\be\label{stat:g_SdS}
	a = b = 1 - \frac{r_{s}}{r} - \frac{\Lambda}{3} r^{2},
\ee
where $r_{s}$ is the Schwarzschild radius encoding the black hole mass and $\Lambda$ a cosmological constant. The values $\Lambda > 0$, $\Lambda = 0$, $\Lambda < 0$ correspond to asymptotically de Sitter (dS), flat and anti-de Sitter (AdS) spacetimes, respectively. Conformal symmetry implies that this Kerr-Schild type static spherically symmetric solution can be slightly generalised by applying \eqref{eq:omega_d0} in which case we obtain\footnote{The even larger of class static spherically symmetric solutions thus reads
\be
	a = \Omega(r)^2 - \frac{r_{s}}{r} \Omega(r)^3 - \frac{\Lambda}{3} r^{2}, \qquad
	b = \frac{\Omega(r)^2 - \frac{r_{s}}{r} \Omega(r)^3 - \frac{\Lambda}{3} r^{2}}{\left(\Omega(r) - r \, \frac{\td \Omega(r)}{\td r} \right)^2},
\ee
where $\Omega(r)$ is an arbitrary function and $r_{s}$ and $\Lambda$ are constants.}
\be\label{eq:Weyl_BH}
	a  = b = s - \frac{r_{s}}{r} + \frac{1 - s^2 }{3r_{s}} r - \frac{\bar \Lambda}{3} r^{2},
\ee
where $s$ is a new free real parameter related to the free parameter in \eqref{eq:omega_d0}. The two cosmological constants are related by $\bar \Lambda = \Lambda + (s-1)^2(s+2)/(9r_{s}^2)$. This is the general static spherically symmetric solution of the Kerr-Schild form in Weyl gravity~\cite{Mannheim:1988dj}.

A large class of wormhole solutions is now readily obtained by applying the Weyl transformation \eqref{stat:g_sss_c} with $\Omega$ specified by \eqref{eq:omega_wh} to \eqref{stat:g_SdS}. We obtain the metric \eqref{stat:g_sss_wh}
with $\hat a$ determined by
\be\label{eqn:geomfunc}
	\hat a = \left(1 + \frac{l^2}{d^2}\right)\left(\eta(l)^2 - \frac{r_{s}}{d}\, \eta(l)^3 -  \frac{d^2\Lambda}{3} \right),
\ee
where $\eta  \equiv  \eta_0 - \atan\left(l/d\right)$ as above. An identical form would have been obtained by applying the transformation to the seemingly more general metric \eqref{eq:Weyl_BH}, because all dependence on $s$ can be absorbed by an appropriate redefinition of $\eta_0$ and $\Lambda$.

For $r_s = 0$ the space is conformally flat and so we might impose $\Lambda = 0$ without loss of generality. We will not do that, however, because first, the conformal transformation turning (A)dS spacetimes into flat ones depend also on time and are thus not covered by the class of conformal transformations considered here. Second, conformal transformations between Einstein spaces are severely restricted, as in most cases they imply constant $\Omega$ or they require the space to have constant curvature\footnote{More precisely, it can be shown that conformal transformations between Einstein spaces should satisfy $(\Omega^{-1})_{; \mu\nu} \propto g_{\mu\nu}$ which implies the integrability condition $C^{\mu}{}_{\nu \rho \sigma} \Omega_{, \mu} = 0$. In particular, if $\Omega$ connects two static spherically symmetric Einstein spaces then $\Omega_{, \mu} \neq 0$ implies that these spaces are locally conformally flat. See also Ref.~\cite{10.2307/2160584}.}. Therefore the S-(A)dS black holes are never locally conformal to Schwarzschild black hole, that is when $r_s \neq 0$ conformal transformations that locally set $\Lambda = 0$ do not exist.

\subsection{Weyl gravity}

Conformal Weyl gravity~\cite{Weyl:1917gp,Weyl1918,Weyl:1918ib,Weyl:1919fi} is a higher derivative theory of gravity described by the action
\be\label{eq:action_Weyl}
	S
	= -\alpha \int \td x^{4} \, \sqrt{-g}  \, C_{\mu \nu \rho \sigma} C^{\mu \nu \rho \sigma},
\ee
where  $C_{\mu \nu \rho \sigma}$ denotes Weyl tensor and $\alpha$ is a dimensionless parameter. The vacuum field equations are the Bach equations
\be\label{eq:eom_Weyl}
	\nabla_{\rho}\nabla_{\sigma}C^{\mu\rho\nu\sigma} + \frac{1}{2}C^{\mu\rho\nu\sigma}R_{\rho\sigma} = 0,
\ee
where $R_{\mu \nu}$ is the Ricci tensor. The symmetries of the Weyl tensor and the second Bianchi identity imply that the Bach equation is solved by Einstein spaces defined by the property
\be\label{eq_einstein_vac}
	R_{\mu\nu} = \Lambda g_{\mu\nu},
\ee
and thus all vacuum solutions of GR with an arbitrary cosmological constant, such as the S-(A)dS spacetimes, are vacuum solutions in Weyl gravity.

The S-(A)dS solutions extended to real coordinates on $\mathbb{R}^{*}$ solve the Bach equation and thus the transformation \eqref{stat:g_sss_wh} can be used in its full extent to obtain new solutions by conformal transformations. Up to conformal transformations the S-(A)dS is also the most general static spherically symmetric solution in Weyl gravity. This follows from the fact that Eq. \eqref{eq:Weyl_BH} is the general Kerr-Schild type spherically symmetric solution to the Bach equation~\cite{Mannheim:1988dj}.

We conclude that in Weyl gravity Eq. \eqref{eqn:geomfunc} describes the general wormhole solution of the form \eqref{stat:g_sss_wh}. This does not mean, however, that other static spherically symmetric wormhole solutions might not be found, e.g., by relaxing the ansatz \eqref{stat:g_sss_wh}. Our primary goal is to demonstrate that the vacuum of Weyl gravity contains a large class of wormholes, implying that they can exist without any need for exotic matter. In a fictional universe in which conformal symmetry is exact these wormholes are as physical as black holes.

\subsection{Conformal scalar-tensor gravity}

Consider scalar-tensor gravity with an action
\begin{align}\label{eq:action_CS}
	S
	= \int \td x^{4} \, \sqrt{-g} \left(\frac{1}{12} \phi^{2} R + \frac{1}{2}(\partial\phi)^{2} - \frac{\lambda}{4} \phi^{4}\right).
\end{align}
This theory is symmetric under the Weyl transformation
\be
	g_{\mu\nu}(x) \mapsto \Omega(x)^{2} g_{\mu\nu}(x), \qquad
	\phi(x) \mapsto \Omega(x)^{-1}\phi(x).
\ee
The scalar is, however, a non-dynamical gauge degree of freedom that may be explicitly removed by choosing $\Omega(x) =  \sqrt{6}\mpl \phi(x)$, which corresponds to the gauge fixing condition $\phi = \sqrt{6}\mpl$. As a result we are left with Einstein gravity with a cosmological constant $\Lambda \equiv 9\lambda \mpl^{2}$:
\be\label{eq:action_CSlambda}
	S =  \int \td x^{4} \, \sqrt{-g} \,\frac{1}{2} \mpl^{2}  (R - 2 \Lambda).
\ee
It now follows that the theory has the S-(A)dS spacetime \eqref{stat:g_SdS} as a solution. Unlike in Weyl gravity, however, $\Lambda$ is not a free parameter, but determined by the strength of the self-interaction $\lambda$ of the scalar. We remark that \eqref{eq:action_CS} is an example of an $f(R,\phi)$ theory, which in specific cases has been shown to contain wormholes without the need for exotic matter~\cite{Zubair:2017oir}.

A wormhole solution is again obtained by applying the Weyl transformation \eqref{eq:omega_wh} to the S-(A)dS metric \eqref{stat:g_SdS}. Now, however, it is accompanied by a non-trivial spherically symmetric scalar field configuration
\be\label{eq:phi}
	\phi(l) =  \frac{\sqrt{6}\mpl}{\sqrt{1+ l^2/d^2}(\eta_0 - \atan\left(l/d\right)) },
\ee
that lives in the wormhole spacetime given by \eqref{eqn:geomfunc}. The field configuration is non-singular for $|\eta_0| \geq \pi/2$ so the wormhole is constructed from the regions of the S-(A)dS spacetime that do not extend through spatial infinity. For a strict inequality $|\eta_0| > \pi/2$ the scalar field behaves asymptotically as $\phi(l) \propto 1/l$ and so it approaches zero at spatial infinity. In case $|\eta_0| = \pi/2$ the field approaches a constant non-zero value at positive or negative infinity. Consider, for example, the case to $\eta_0 = \pi/2$: the wormhole is obtained from the patch $ r \in (r_{-}, \infty)$ so that $r \to \infty$ corresponds to $l \to \infty$, in which case the field approaches a constant non-zero value, and $r \to r_{-}$ corresponds to $l \to -\infty$ and the field tends to zero.

A vanishing value of the field corresponds to an infinite effective gravitational coupling in conformal scalar-tensor gravity \eqref{eq:action_CS}. This seems to be an inevitable feature of traversable vacuum wormholes in more general scalar-tensor theories, because traversable wormholes and an effective gravitational coupling that is everywhere non-vanishing and finite are possible only in presence of matter violating the null energy-condition~\cite{Butcher:2015sea}. Nevertheless, we stress that $\phi \to 0$ is approached only asymptotically - gravity may become stronger as we travel farther from the wormhole, but never actually infinite for any finite distance $l$. If one side of the wormhole is to represent the observable universe, then it is necessary to impose $|\eta_0| = \pi/2$. The other side of such wormholes will still exhibit an unbounded growth of the gravitational force as the distance increases and will thus be qualitatively very different from the observable universe. 

Observe that the scalar field here is a gravitational degree of freedom and no matter fields are present. Thus the null energy condition $T_{\mu\nu}N^{\mu}N^{\nu} \geq 0$, where $N^{\mu}$ are null vectors and $T_{\mu\nu}$ is the usual Hilbert stress-energy-momentum tensor, is trivially satisfied since $T_{\mu\nu} = 0$. To develop a better intuition between the null energy condition for \eqref{eq:action_CS} and for general relativity, it is instructive to express the field equations of the action \eqref{eq:action_CS} as $G_{\mu\nu} = Q_{\mu\nu}$, where $Q_{\mu\nu}$ are the terms involving the scalar field and its derivatives. One then realises that the scalar field mimics the behaviour of exotic matter violating the null energy condition, since for the wormhole spacetime \eqref{eqn:geomfunc} there exist null-vectors $N$ which satisfy $R_{\mu\nu}N^{\mu}N^{\nu} < 0$ and so $Q_{\mu\nu}N^\mu N^\nu<0$. Thus in conformal scalar-tensor gravity, the wormhole is present without exotic dark matter but is sourced by the gravitational scalar.

\subsection{Classification and examples}
\label{ssec:examples}
Having the general wormhole solution~\eqref{stat:g_sss_wh} at hand we now study the geometric properties which characterise the solution. We focus on the traversability and asymptotics of the spacetimes connected by the wormhole.  These properties can be used as a basis of a qualitative classification of the wormhole solutions. A few simple examples are discussed in detail in the following subsections.

Many features of the wormhole can be intuitively understood from the fact that null geodesics are invariant under conformal transformations. As the metric \eqref{stat:g_sss_wh}  is conformally related to a region of the S-(A)dS metric, its null geodesics match the null geodesics of the S-(A)dS spacetime up to reparametrization. The geodesics of the S-(A)dS have been discussed in detail in Ref.~\cite{Hackmann:2008zz}. 

Conformal invariance of the causal structure implies that the wormhole will inherit its horizons from the region of the extended S-(A)dS spacetime \eqref{stat:g_SdS} it is conformally related to. Especially, if $r_s  = 0$ the wormhole is formed from a pure dS or AdS spacetime, i.e., a spacetime without a black hole horizon, and the condition $\Lambda \leq 0$ implies the absence of a dS horizon. Traversability can be loosely rephrased as: the wormhole throat should not be surrounded by horizons.  This does, however, not imply that the spacetime does not contain any horizons.

The general asymptotics can be obtained by expanding \eqref{eqn:geomfunc}. The first four terms read
\begin{align}\label{hat_a_expansion}
	\hat a(l/d \to \pm \infty)
&	= - \frac{1}{3}\Lambda_{\pm} l^{2}
	+ \mu_{\pm} l
	+ s_{\pm}
	- \frac{r_{s\pm}}{l}
	+ \mathcal{O}\left(l \right)^{-2} ,
\end{align}
where
\begin{align}\label{WH_asymp}
	\Lambda_{\pm}
&	= \Lambda + 3\frac{r_{s} -  r_{\pm}}{ r_{\pm}^{3}}
	, \nonumber\\
	\mu_{\pm}
&	=	\frac{2 r_{\pm} - 3r_s}{ r_{\pm}^2}
	, \nonumber\\
	s_{\pm}
&	=   1 - \frac{3r_{s}}{ r_{\pm}} - \frac{d^2}{3} \Lambda_{\pm}
	, \nonumber\\
	r_{s\pm}
&	= r_s - \frac{2 d^2 }{3} \mu_{\pm} ,
\end{align}
and $ r_{\pm}$ denote the values of the radial coordinate at the boundary of the patch of the initial S-(A)dS spacetime from which the wormhole is constructed. Note that $d$ and $ r_{\pm}$ are related by Eq.~\eqref{r_pm_d_eta0}.  As already implied by~\eqref{hat_a_expansion}, the wormhole connects two spacetimes with an asymptotically constant curvature, so that
\be
	R_{\mu\nu\rho\sigma} \sim \frac{\Lambda_{\pm}}{3} (g_{\mu\rho}g_{\nu\sigma} - g_{\mu\sigma}g_{\nu\rho}) + \mathcal{O} (l^{-1}),
\ee
for $l/d \to \pm \infty$, respectively. The $l^{-1}$ term is proportional to $\mu_{\pm}$ and thus the Riemann curvature is constant roughly when
\be\label{mu_lambda}
	l \lesssim-\left| \mu_{-}/\Lambda_{-} \right| \qquad \mbox{or} \qquad
	l \gtrsim \left|\mu_{+}/\Lambda_{+}\right|.
\ee
Of course, for vanishing $\mu_{+}$ one needs to look at the higher order terms in the expansion~\eqref{hat_a_expansion}.

It is not possible to find solutions for which both $\Lambda_{\pm}$ and both $\mu_{\pm}$ vanish, so the condition \eqref{mu_lambda} implies that this class of solutions does not contain wormholes connecting asymptotically flat spacetimes. Apart form a brute-force calculation, this can be seen by first noting that the wormhole, as determined by \eqref{eqn:geomfunc}, is described by 4 parameters, e.g., $\Lambda$, $r_s$, $ r_{\pm}$.\footnote{An alternative parametrisation replacing $ r_{\pm}$ with $\eta_0$ and $d$  can be obtained by \eqref{r_pm_d_eta0}.} One of the parameters, however, can be fixed by choosing the units, so up to a rescaling the metric is determined by only 3 parameters and thus there is not enough freedom to choose $\Lambda_{\pm} = 0$ and $\mu_{\pm} = 0$ simultaneously.

Nevertheless, it is possible to freely choose the first four parameters of the asymptotic expansion~\eqref{WH_asymp} on one side of the wormhole and this choice will completely determine the asymptotic properties of the other side of the wormhole. For example, consider the case where the $l > 0$ side asymptotically resembles a black hole in the observable universe, for which $\mu_{+} = 0$. The condition $\mu_{+} = 0$ corresponds to $r_{+} = \infty$ or to $r_{+} = 3/2 r_s$. If we choose the first option, the $l > 0$ side of the wormhole inherits the asymptotics of the original S-(A)dS spacetime, for which we assume that the cosmological horizon is much larger than other involved scales, $\Lambda^{-1/2} \gg d, r_s$. It is now possible to choose $ r_{-}$ so that $\mu_{-} = 0$ or $\Lambda_{-} = 0$. In both cases the relevant length scales $\Lambda_{-}^{-1/2}$ or $\mu_{-}^{-1}$ would be comparable to the throat radius, $d =  r_{-}/\pi$. Consequently, it is not only impossible to set $\Lambda_{\pm} = 0$, $\mu_{\pm} = 0$ on both sides of the wormhole, but doing so on one side of the wormhole implies that the other side does not resemble an asymptotically S-(A)dS black hole. By tuning the parameters so that $\Lambda_{-}$ is small enough the $l<0$ side will be approximately (A)dS for $\left|\mu_{+}/\Lambda_{+}\right| \gg l$. This is discussed further in Section~\ref{ssec:ex_dSdS}.

Finally, it is instructive to compare this expansion to the Weyl gravity black hole \eqref{eq:Weyl_BH} which has a similar structure to the asymptotic wormhole exteriors.  The black hole \eqref{eq:Weyl_BH} would, however, satisfy $s_{\pm}^2 + 3 r_{s\pm} \mu_{s\pm} = 1$, thus not all wormhole exteriors can be asymptotically matched to the black hole solution \eqref{eq:Weyl_BH}.

\subsubsection{Traversable AdS/Minkowski wormhole}
\label{ssec:adsflat}

\tikzsetfigurename{penrose_}
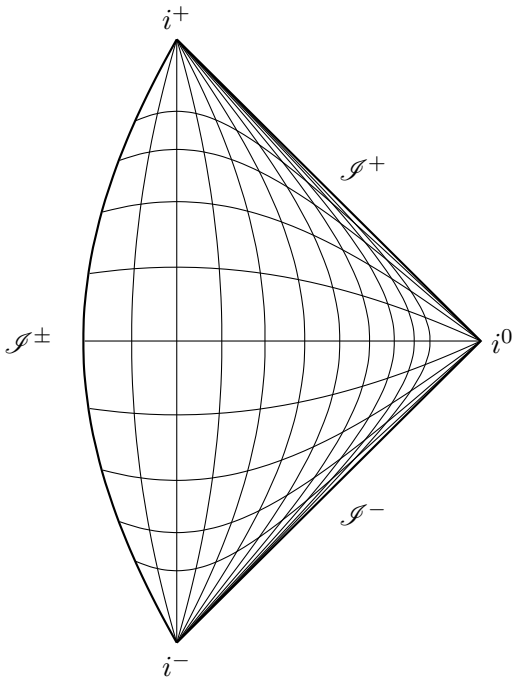
\begin{figure}[htp]
	\centering
	\begin{tikzpicture}
	\draw[domain=-5:5,smooth,variable=\x,samples=50,thick] plot ({2*tanh(\x-0.3182) - 2*tanh(\x+0.3182)},{2*tanh(\x-0.3182) + 2*tanh(\x+0.3182)});
	\draw[thick] (0,-4) -- (4,0) -- (0,4);
	\foreach \y in {-0.15,0.0,...,1.2} \draw[domain=-5:5,smooth,variable=\x,samples=50] plot ({2*tanh(\x+\y) - 2*tanh(\x-\y)},{2*tanh(\x+\y) + 2*tanh(\x-\y)});
	\foreach \y in {-1.0,-0.75,...,1.0} \draw[domain=-0.3128:5,smooth,variable=\x,samples=50] plot ({2*tanh(\y+\x) - 2*tanh(\y-\x)},{2*tanh(\y+\x) + 2*tanh(\y-\x)});
	\draw (0,4) node[above] {$i^+$};
	\draw (0,-4) node[below] {$i^-$};
	\draw (4,0) node[right] {$i^0$};
	\draw (2,2) node[above right] {$\mathscr{I}^+$};
	\draw (2,-2) node[below right] {$\mathscr{I}^-$};
	\draw (-1.5,0) node[left] {$\mathscr{I}^{\pm}$};
	\end{tikzpicture}
	\caption{Penrose diagram for the AdS/Minkowski wormhole. Horizontal lines show hypersurfaces with constant time $t$, while vertical lines have constant radial coordinate $l$. The vertical line in the center corresponds to the throat $l = 0$. On the left is the AdS part, where both future and past lightlike geodesics reach the conformal boundary $\mathscr{I}^{\pm}$ in finite coordinate time $t$. On the right is the asymptotically flat part, with the usual boundary components of Minkowski space.}
	\label{fig:asympen}
\end{figure}

Wormholes connecting an AdS spacetime to a Minkowski spacetime can be constructed from an asymptotically flat spacetime, in the simplest case the Minkowski spacetime. In order to inherit the asymptotic flatness we need to send $ r_{+}$ to infinity (or $ r_{-}$ to minus infinity). In all, one of the simplest examples of such a spacetime corresponds to $\Lambda = 0$, $r_s = 0$, $\eta_0 = (-)\pi/2$ with the throat radius $d$ as the only free parameter. The metric then reads
\be\label{g_asympen}
	\td s^2 =	- \left(\atan \frac{l}{d} \pm \frac{\pi}{2}\right)^2  \left(1 + \frac{l^2}{d^2}\right)\,\td t^2 \,  + \frac{\td l^2}{\left(\atan \frac{l}{d} \pm \frac{\pi}{2}\right)^2  \left(1 + \frac{l^2}{d^2}\right)} \,  + (d^2+l^2) \td  \Omega^2.
\ee
This wormhole connects an asymptotically flat spacetime to an asymptotically AdS spacetime with $\Lambda = -\pi^2/d^2$.

The causal structure of this wormhole is illustrated by the Penrose diagram in Fig~\ref{fig:asympen}. The coordinates used are given in Appendix~\ref{ssec:penrose}. For a more quantitative understanding consider radially falling lightlike geodesics for \eqref{g_asympen},
\be
	\gamma(t) = \left( t, d \cot\left(\frac{1}{2/\pi + t/d} \right), \frac{\pi}{2}, 0 \right),
\ee
where the constant of integration was chosen so that, $l(0) = 0$. The coordinate time is seen to be bounded, $t-t_0 \in \left(-d/\pi, \infty\right)$, so light rays traverse the AdS region $l \in (-\infty, 0)$ in finite coordinate time $\Delta t = d/\pi$. This is characteristic for asymptotically AdS spacetimes. Thus an observer near the throat of the wormhole sees a photon coming from $l=-\infty$ reaching its position in finite time and then disappearing to $l=\infty$ in infinite time.

\subsubsection{Traversable AdS/AdS wormholes}
\label{ssec:adssym}

\tikzsetfigurename{penrose_}
\begin{figure}[htp]
	\centering
	\begin{tikzpicture}
	\draw[domain=-5:5,smooth,variable=\x,samples=50,thick] plot ({2*tanh(\x-1) - 2*tanh(\x+1)},{2*tanh(\x-1) + 2*tanh(\x+1)});
	\draw[domain=-5:5,smooth,variable=\x,samples=50,thick] plot ({2*tanh(\x+1) - 2*tanh(\x-1)},{2*tanh(\x+1) + 2*tanh(\x-1)});
	\foreach \y in {-0.8,-0.6,...,0.8} \draw[domain=-5:5,smooth,variable=\x,samples=50] plot ({2*tanh(\x-\y) - 2*tanh(\x+\y)},{2*tanh(\x-\y) + 2*tanh(\x+\y)});
	\foreach \y in {-1.0,-0.75,...,1.0} \draw[domain=-1:1,smooth,variable=\x,samples=50] plot ({2*tanh(\y-\x) - 2*tanh(\y+\x)},{2*tanh(\y-\x) + 2*tanh(\y+\x)});
	\draw (0,4) node[above] {$i^+$};
	\draw (0,-4) node[below] {$i^-$};
	\draw (3,0) node[right] {$\mathscr{I}^{\pm}$};
	\draw (-3,0) node[left] {$\mathscr{I}^{\pm}$};
	\end{tikzpicture}
	\caption{Penrose diagram for the symmetric AdS/AdS wormhole. The causal structure of the spacetime agrees with two copies of the left half of the spacetime shown in figure~\ref{fig:asympen}, i.e., the Minkowski part has been replaced with another AdS part.}
	\label{fig:sympen}
\end{figure}
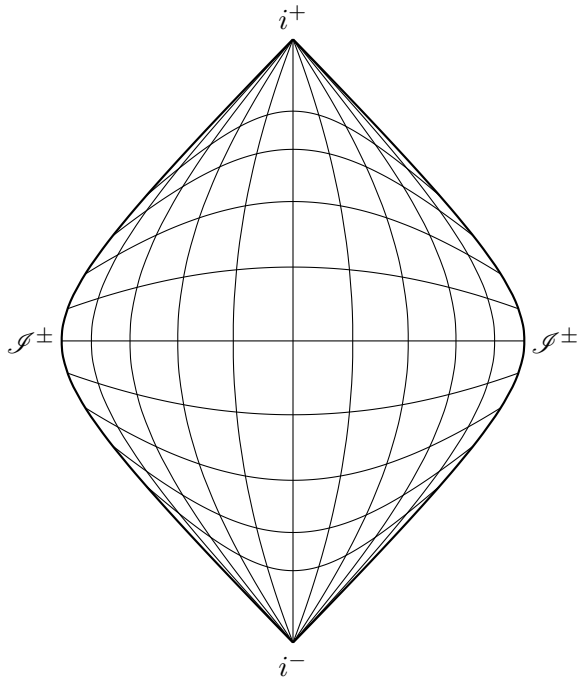

The leading term $l^2$ in the expansion \eqref{hat_a_expansion} is characteristic for (A)dS spacetimes. By Eq.~\eqref{hat_a} it is also proportional to the function $a$ of original the S-(A)dS spacetime \eqref{stat:g_SdS}. Therefore, as we transform the S-(A)dS black hole into a wormhole, the spatial boundary of the restriction is transformed an asymptotically (A)dS spacetime unless the restriction is made at a horizon. If the restriction is made between the black hole and cosmological horizons, the resulting wormhole will connect two AdS spacetimes.

If the original spacetime is asymptotically AdS, it is possible to construct a traversable wormhole from the hypertube defined by $ r \in \mathbb{R}^{*}\setminus[ r_{+}, r_{-}]$ of the extended spacetime provided that it does not contain any horizons. This is the example we will consider next. For simplicity we choose $r_s = 0$ and $\eta_0 = 0$ implying that $r_{-} = -r_{+} = 2d/\pi$. The wormhole throat is positioned at the spatial infinity of the original spacetime. The metric reads
\be\label{g_symp}
	\td s^2
	=	- \left(\atan^2 \frac{l}{d} + \frac{d^2}{l_\Lambda^2}\right) \left(1 + \frac{l^2}{d^2}\right)\,\td t^2 \,  + \frac{\td l^2}{\left(\atan^2 \frac{l}{d} + \frac{d^2}{l_\Lambda^2}\right)  \left(1 + \frac{l^2}{d^2}\right)} \,  + (d^2+l^2) \td  \Omega^2,
\ee
where we defined $l_\Lambda \equiv \sqrt{-3/\Lambda} > 0$. In case $l_\Lambda^{2} < 0$, i.e., if the original spacetime would be dS, the wormhole would not be traversable. By Eq.~\eqref{WH_asymp}, the wormhole connects two asymptotically AdS spacetimes with
\be
	\Lambda_{\pm} = -\frac{3}{d^2}\left(\frac{\pi^2}{4} + \frac{d^2}{l_\Lambda^2}\right),
\ee
so the distance scales associated with the asymptotic curvature are comparable to or smaller than the throat size.

The causal structure of this wormhole is illustrated by the Penrose diagram in Fig.~\ref{fig:sympen}. Consider again radially falling lightlike geodesics of \eqref{g_asympen},
\be
	\gamma(t) = \left( t,d \tan\left(\frac{d}{l_\Lambda} \tan\left(\frac{t}{l_\Lambda}\right)\right) , \frac{\pi}{2}, 0 \right),
\ee
the coordinate time is seen to be bounded, $t \in ( - l_\Lambda \atan\left(\pi l_\Lambda/2d\right), l_\Lambda \atan\left(\pi l_\Lambda/2d\right) )$. As expected, the light rays traverse the AdS regions $l \in (\pm\infty, 0)$ in finite coordinate time $\Delta t = l_\Lambda \atan\left(\pi l_\Lambda/2d\right)$, so an observer near the throat of the wormhole sees that photons travelling through the wormhole take a finite time to travel from $l = -\infty$ to $l = +\infty$.

Although the metric \eqref{g_symp} was obtained from an AdS spacetime it could as well be obtained by conformally restricting/expanding a Minkowski space, which is conformal to AdS. We stress again that the conformality between this wormhole to AdS is only local. Moreover, it is possible to spatially compactify the wormhole spacetime by gluing together the negative and positive spatial infinities, so that $l$ takes values on $\mathbb{R}^{*}$, resulting in a spacetime that is spatially compact and locally conformally flat. This spacetime can not be conformally extended to an AdS spacetime. In conclusion, both wormholes \eqref{g_symp} and \eqref{g_asympen} in the two examples discussed above  are solutions of any conformal theory that has the Minkowski spacetime as a solution.

\subsubsection{Traversable dS/dS wormholes}
\label{ssec:ex_dSdS}

\begin{figure}[htp]
	\centering
	\begin{tikzpicture}[scale=0.9]
	\draw[thick] (-8,4) -- (8,4) -- (8,-4) -- (-8,-4) -- (-8,4);
	\draw[thick] (-8,4) -- (0,-4) -- (8,4);
	\draw[thick] (-8,-4) -- (0,4) -- (8,-4);
	\foreach \y in {-1.0,-0.75,...,1.0} \draw[domain=-5:5,smooth,variable=\x,samples=50,dashed] plot ({2*tanh(\x+\y) - 2*tanh(\x-\y)},{2*tanh(\x+\y) + 2*tanh(\x-\y)});
	\foreach \y in {-1.0,-0.75,...,1.0} \draw[domain=-5:5,smooth,variable=\x,samples=50] plot ({2*tanh(\y+\x) - 2*tanh(\y-\x)},{2*tanh(\y+\x) + 2*tanh(\y-\x)});
	\foreach \y in {-1.0,-0.75,...,0.0} \draw[domain=-5:5,smooth,variable=\x,samples=50,dashed] plot ({2*tanh(\x+\y) - 2*tanh(\x-\y) + 8},{2*tanh(\x+\y) + 2*tanh(\x-\y)});
	\foreach \y in {0.0,0.25,...,1.0} \draw[domain=-5:5,smooth,variable=\x,samples=50,dashed] plot ({2*tanh(\x+\y) - 2*tanh(\x-\y) - 8},{2*tanh(\x+\y) + 2*tanh(\x-\y)});
	\foreach \y in {-1.0,-0.75,...,1.0} \draw[domain=-5:0,smooth,variable=\x,samples=50] plot ({2*tanh(\y+\x) - 2*tanh(\y-\x) + 8},{2*tanh(\y+\x) + 2*tanh(\y-\x)});
	\foreach \y in {-1.0,-0.75,...,1.0} \draw[domain=0:5,smooth,variable=\x,samples=50] plot ({2*tanh(\y+\x) - 2*tanh(\y-\x) - 8},{2*tanh(\y+\x) + 2*tanh(\y-\x)});
	\foreach \y in {-1.0,-0.75,...,1.0} \draw[domain=-5:0,smooth,variable=\x,samples=50] plot ({2*tanh(\x+\y) - 2*tanh(\x-\y) + 4},{2*tanh(\x+\y) + 2*tanh(\x-\y) + 4});
	\foreach \y in {-1.0,-0.75,...,1.0} \draw[domain=0:5,smooth,variable=\x,samples=50] plot ({2*tanh(\x+\y) - 2*tanh(\x-\y) + 4},{2*tanh(\x+\y) + 2*tanh(\x-\y) - 4});
	\foreach \y in {-1.0,-0.75,...,1.0} \draw[domain=-5:0,smooth,variable=\x,samples=50] plot ({2*tanh(\x+\y) - 2*tanh(\x-\y) - 4},{2*tanh(\x+\y) + 2*tanh(\x-\y) + 4});
	\foreach \y in {-1.0,-0.75,...,1.0} \draw[domain=0:5,smooth,variable=\x,samples=50] plot ({2*tanh(\x+\y) - 2*tanh(\x-\y) - 4},{2*tanh(\x+\y) + 2*tanh(\x-\y) - 4});
	\foreach \y in {-1.0,-0.75,...,0.0} \draw[domain=-5:5,smooth,variable=\x,samples=50,dashed] plot ({2*tanh(\y+\x) - 2*tanh(\y-\x) + 4},{2*tanh(\y+\x) + 2*tanh(\y-\x) + 4});
	\foreach \y in {0.0,0.25,...,1.0} \draw[domain=-5:5,smooth,variable=\x,samples=50,dashed] plot ({2*tanh(\y+\x) - 2*tanh(\y-\x) + 4},{2*tanh(\y+\x) + 2*tanh(\y-\x) - 4});
	\foreach \y in {-1.0,-0.75,...,0.0} \draw[domain=-5:5,smooth,variable=\x,samples=50,dashed] plot ({2*tanh(\y+\x) - 2*tanh(\y-\x) - 4},{2*tanh(\y+\x) + 2*tanh(\y-\x) + 4});
	\foreach \y in {0.0,0.25,...,1.0} \draw[domain=-5:5,smooth,variable=\x,samples=50,dashed] plot ({2*tanh(\y+\x) - 2*tanh(\y-\x) - 4},{2*tanh(\y+\x) + 2*tanh(\y-\x) - 4});
	\draw (4,4) node[above right] {$\mathscr{I}^+$};
	\draw (4,-4) node[below right] {$\mathscr{I}^-$};
	\draw (-4,4) node[above right] {$\mathscr{I}^+$};
	\draw (-4,-4) node[below right] {$\mathscr{I}^-$};
	\end{tikzpicture}
	\caption{Penrose diagram for the dS-dS wormhole. Solid lines show hypersurfaces with constant time $t$, while dashed lines have constant radial coordinate $l$. The central patch of the diagram contains the throat of the wormhole, which is located at the vertical line in the center. On both sides of the wormhole there exist cosmological de Sitter horizons, indicated by diagonal lines, which prevent certain regions of the asymptotic spacetimes to be reached from the throat, and vice versa. Regions on the far left and right of the diagram are causally disconnected from the throat in the center.}
	\label{fig:dsds}
\end{figure}
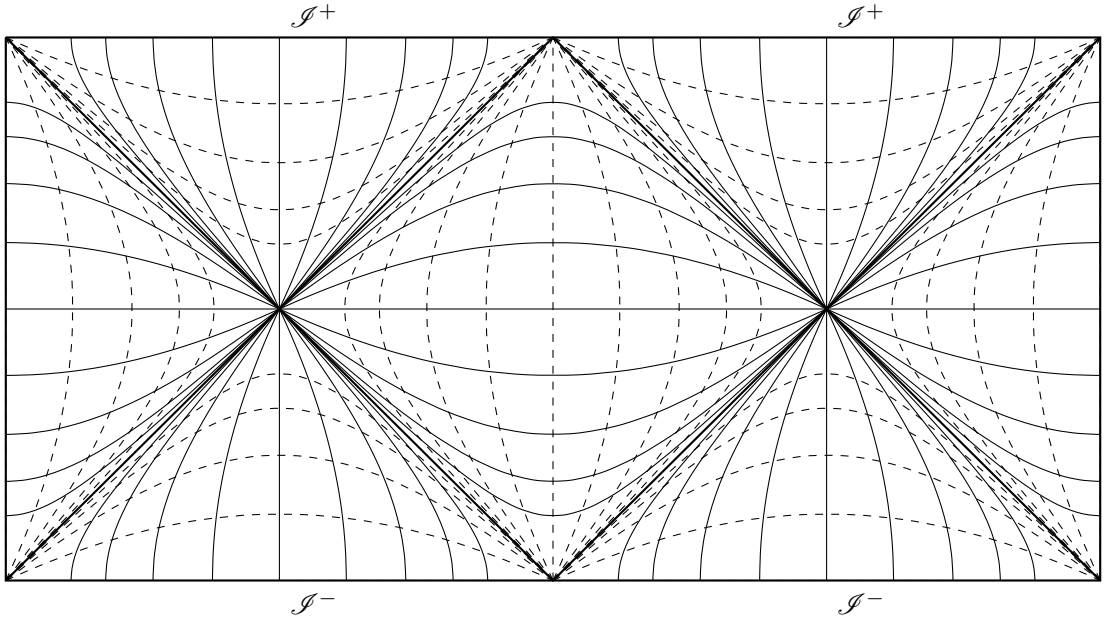

Unlike for the previous examples, wormholes connecting asymptotically dS spacetimes can not be constructed from flat spacetimes. As the horizons are inherited from the original spacetime it is necessary that $\Lambda, r_s > 0$. We will work with the region $ r \in ( r_{-}, \infty)$, i.e., we send $ r_{+}$ to infinity. To obtain a second dS horizon the cut at $ r_{-}$ has to be made slightly within the event horizon of the black hole. By choosing  $ r_{-} = r_{s}$ we obtain by using \eqref{WH_asymp} that
\begin{align}\label{a_dSdS}
	\hat a \approx
	\left\{
	\begin{array}{ll}
	1 - \frac{r_{s}}{l} - \frac{\Lambda}{3} l^2, & \mbox{ when } l \to \infty ,\\
	-2 - \frac{r_{s}}{l} - \frac{l}{r_{s}} - \frac{\Lambda}{3} l^2, & \mbox{ when } l \to -\infty .
	\end{array}
	\right.
\end{align}
From the leading terms in the expansion it can be seen that both sides of the wormhole possess cosmological horizons at  $l_{H+} \approx \sqrt{3/\Lambda}$ and $l_{H-} \approx -3/(r_{s}\Lambda)$. $\hat a$ is positive for $l \in (l_{H-},l_{H+})$ so these are also the only horizons. It is clear that the $l > 0$ side of the wormhole is asymptotically a S-dS black hole. The expansion \eqref{a_dSdS} of the $l<0$ side may also look asymptotically dS, yet within the horizon the linear term dominates and the $l^2$ term becomes dominant only beyond the horizon, when $l \ll l_{H-}$.

The Penrose diagram of the dS/dS wormhole is depicted in Fig.~\ref{fig:dsds}. The throat is the central vertical line and corresponds to \(l = 0\). For an observer at rest at the throat, the spacetime decomposes into different regions, which are separated by a number of horizons. The two diagonals which meet at \(\mathscr{I}_+\) above the throat mark the event horizons. Any events outside of these horizons cannot be seen by the observer, and hence no information from there can pass through the wormhole. These are complemented by two particle horizons denoted by the diagonal lines meeting at \(\mathscr{I}_-\) below the throat. No particles or information coming through the wormhole can reach these horizons. In particular, the two triangular regions at the edges of the Penrose diagram are completely causally disconnected from the wormhole throat, and thus an observer located in these regions cannot interact with the wormhole in any way.

\subsection{Inclusion of matter: Charged wormholes}

Our construction can be straightforwardly generalised to conformal theories containing matter fields. As before, without loss of generality it is sufficient to study spherically symmetric solutions satisfying the Kerr-Schild ansatz for the metric, Eq.~\eqref{stat:g_sss} with $a = b$. Having such solutions at hand it is then possible to apply the wormhole transformation~\eqref{hat_a} to obtain wormhole solutions that are locally conformal to the original spherically symmetric spacetime. In that case, the Weyl transformation should also be applied to the matter fields.

The wormhole geometries given by \eqref{g_symp} and \eqref{g_asympen} exist in any conformal theory that has the Minkowski spacetime as a solution. It is reasonable to expect that the latter property is shared by most gravitational theories, so these (vacuum) wormhole geometries are nearly universal. As gravitational interactions of matter fields will generally differ in different conformal theories of gravity, this universality is lost for non-vacuum solutions.

The static spherically symmetric vacuum solutions of Weyl gravity contain the vacuum solutions of conformal scalar-tensor gravity. It is interesting to consider the differences between non-vacuum solutions of these theories. A simple example of conformal matter fields is the electromagnetic field, described by the action
\be\label{S_EM}
	S
	= - \frac{1}{4} \int \td x^{4} \, \sqrt{-g}  \, F_{\mu\nu} F^{\mu\nu},
\ee
where $F_{\mu\nu} = \partial_{\mu}A_{\nu} - \partial_{\nu}A_{\mu}$ is the electromagnetic field tensor and $A_{\mu}$ the vector potential. Note that the action~\eqref{S_EM} is invariant under Weyl transformations, provided that the vector potential is left unchanged. The Maxwell equations are therefore also invariant. For static spherically symmetric field configurations on the background Eq.~\eqref{stat:g_sss} they imply
\be\label{A_EM}
	A_{\mu} = (-\sqrt{2}Q/r, 0, 0, 0),
\ee
where $Q$ is the electric charge~\cite{Oda:2016nvc}. Consider now the static spherically symmetric solutions for Weyl gravity and conformal scalar-tensor gravity:
\begin{itemize}
\item In conformal scalar-tensor gravity the Kerr-Schild type solution can be obtained in the Einstein gauge $\phi = \sqrt{6}\mpl$, for which the action for gravity is given by \eqref{eq:action_CSlambda}. The solution is given by the Reissner–Nordstr\"om-(anti) de Sitter metric\footnote{For a more detailed derivation of the Reissner-Nordstr\"om metric in scalar-tensor gravity see for example~\cite{Oda:2016nvc}.} for which,
	\be
		a = 1 - \frac{r_{s}}{r} + \frac{r_{q}^2}{r^2} - \frac{\Lambda}{3} r^{2},
	\ee
	where $r_{q} \equiv Q^2/\mpl^2$. The wormhole geometry obtained from~\eqref{hat_a} reads,
	\be
		\hat a = \left(1 + \frac{l^2}{d^2}\right)\left(\eta(l)^2 - \frac{r_{s}}{d}\, \eta(l)^3 + \frac{r_{q}^2}{d^2}\, \eta(l)^4 -  \frac{d^2\Lambda}{3} \right).
	\ee
	with the conformal scalar given by \eqref{eq:phi} and the electromagnetic potential\footnote{Note that, in general, when applying the conformal transformation \eqref{stat:g_sss_c} relating spherically symmetric metrics, the expression for the electromagnetic vector potential $A_{\mu}$ is affected only by the coordinate transformation $r \to r \Omega$.}
	\be\label{A_EM_WH}
		A_{\mu} = (\sqrt{2} Q \atan\left(l/d\right)/d , 0, 0, 0),
	\ee
	where we neglected the unphysical constant terms. The electromagnetic field is strongest around the throat of the wormhole and decays as $l^{-2}$ at spatial infinity. The different sides of the wormhole have opposite apparent charges as measured by an outside observer.

	\item In Weyl gravity a Kerr-Schild type solution can be obtained by making the ansatz
	\be
		a = s - \frac{r_{s}}{r} + \mu r - \frac{\Lambda}{3} r^{2},
	\ee
	which, after plugging it into the Bach equation with an electromagnetic source,
	\be
		\nabla_{\rho}\nabla_{\sigma}C^{\mu\rho\nu\sigma} + \frac{1}{2}C^{\mu\rho\nu\sigma}R_{\rho\sigma} = -\frac{1}{\alpha}T_{EM}^{\mu\nu},
	\ee
	yields the constraint,
	\be
		1 - s^2 - 3 r_s \mu = \frac{3 Q^2}{2\alpha}.
	\ee
	For $Q = 0$ the solution reduces to the vacuum solution \eqref{eq:Weyl_BH}, as expected. As in the vacuum case, $s$ can be absorbed by a suitable redefinition of the parameters and the metric for the wormhole can be expressed as
	\be
		\hat a = \left(1 + \frac{l^2}{d^2}\right)\left(\eta(l)^2 - \frac{r_{s}}{d}\, \eta(l)^3  - \frac{Q^2 d}{2\alpha r_s} \, \eta(l) -  \frac{d^2\Lambda}{3} \right).
	\ee
	while the electromagnetic vector potential is given by \eqref{A_EM_WH}.

	Unlike in conformal scalar-tensor gravity, inclusion of an electric charge does not  qualitatively change the geometry of the black hole. Quantitatively, the changes can be felt by a modified relation between the parameters $\mu$  and $r_s$. Interestingly, it follows that the asymptotic geometry \eqref{hat_a_expansion} of wormholes can be matched with a charged black hole.

\end{itemize}

\section{Conclusions}
\label{sec:end}

In this paper we demonstrated that conformal theories of gravity contain traversable wormholes if they have any spherically symmetric solution at all, which is a very mild assumption and certainly the case for the conformal theories of gravity of interest. To construct the wormhole solution we started with a spherical symmetric spacetime and applied the conformal transformation which turns subsets of these spacetimes into wormhole spacetimes.

We constructed the wormhole solutions explicitly for two most studied conformal gravity theories, the Weyl gravity and the scalar-tensor gravity with a conformally coupled scalar. Both theories have in common that the extended Schwarzschild - (anti) de Sitter spacetime is a solution of their field equations. We could identify certain subsets of these spacetimes to which we applied a conformal transformation such that they become wormhole spacetimes in which the wormhole connects spaces with asymptotically constant curvature. Finally, we presented three specific examples of traversable wormholes connecting asymptotically AdS/Minkowski, AdS/AdS and dS/dS spacetimes which are all obtained by applying a conformal transformation to a subset of the original Schwarzschild - (anti) de Sitter spacetime. We briefly discuss the inclusion of matter fields and give explicit examples of wormholes surrounded by an electromagnetic field in Weyl gravity and conformal scalar-tensor gravity.

An important feature of these wormholes is that they are already present in the vacuum of the conformal gravity theories considered. Thus among static spherically symmetric spacetimes there is no reason the prefer black holes over wormholes in these theories. Moreover, the wormholes found here may be used as a starting point for a perturbative study in theories where conformal symmetry is only approximate, that is the terms breaking the conformal symmetry are small. This may be the case, for example, in agravity, that was shown to be asymptotically conformal. If the conformal symmetry turns out to be a good symmetry of Nature at small scales, the singularity of a black hole might be removed and replaced by the throat of a wormhole. Moreover, as traversable wormholes can mimic black holes, the throat might even replace the horizon of the gravitating body.

The method presented here can be extended to construct wormholes with more complicated topologies, e.g., by considering initial spacetimes with a nontrivial topology. 

 The main message we aim to convey in this paper is that the traversable wormholes are much less exotic solutions in conformal gravity than previously believed since their existence does not require the existence of exotic matter or negative energy objects. Being a vacuum solution, the wormholes may play a fundamental role in the formulation of an ultraviolet complete theory of gravity as well as provide observable phenomenological consequences beyond general relativity.


\section*{Acknowledgments}
We thank S. Solodukhin for the discussions on the radiation of wormholes.
This work was supported by the Estonian Research Council grants PUT790 and PUT799, via the Mobilitas Plus grant MOBTT5, the grant IUT23-6 of the Estonian Ministry of Education and Research, and by the EU through the European Regional Development Fund CoE program grant TK133 ``The Dark Side of the Universe'', as well as trough the COST Action CANTATA, supported by COST (European Cooperation in Science and Technology).

\appendix

\section{Some properties of the wormhole metric}
\label{A}

In this Appendix we briefly summarise some of the basic features of the wormhole metric \eqref{stat:g_sss_wh} which we repeat here for the convenience,
\be
	\hat g =  \Omega^2  g = - \hat a(l) \, \td t^2 + \hat a(l)^{-1} \td l^2 + (l^2 + d^2) \td \mathbf{\Sigma}^2.
\ee

\subsection{Wormhole throat in the Morris-Thorne formalism }
\label{ssec:throat}

We now discuss the throat of the wormhole in more detail. For this purpose it is convenient to use the coordinates $(t,r,\theta,\phi,...)$, where $r^2 = l^2 + d^2$ is introduced earlier, and to bring the metric to the standard Morris-Thorne form~\cite{Morris:1988cz,Morris:1988tu}
\be
	\td s^2 = -e^{2\Phi_{\pm}(r)}\td t^2 + \frac{\td r^2}{1 - b_{\pm}(r) / r} + r^2 \td \mathbf{\Sigma}^2 \,,
\ee
where the redshift function $\Phi(r)$ and shape function $b(r)$ are given by
\be
	\Phi_{\pm}(r) = \frac{1}{2}\ln\hat{a}_{\pm}(r)\,, \quad
	b_{\pm}(r) = r - \frac{r^2 - d^2}{r}\hat{a}_{\pm}(r)\,,
\ee
where
\be
	\hat{a}_{\pm}(r) = \frac{r^2}{d^2}\left(\frac{\Lambda d^2}{3} + \tilde{\eta}_{\pm}^2(r) - \frac{r_s}{d}\hat{\eta}_{\pm}^3(r)\right)\,, \quad
	\eta_{\pm}(r) = \pm\acos\left(\frac{d}{r}\right) + \eta_0\,.
\ee
Note that these coordinates actually describe two coordinate patches, each of them covering the range $r \in [d,\infty)$. A number of conditions must be satisfied in order for the solution to describe the shape of a wormhole, which we discuss in the following.

\begin{itemize}
\item
The time coordinate must be continuous at the throat, which implies that $\Phi_+(d) = \Phi_-(d)$. This is the case, since one easily checks that
\be
	\eta_+(d) = \eta_0 = \eta_-(d)\,,
\ee
and hence
\be
	\hat{a}_+(d) = \frac{\Lambda d^2}{3} + \eta_0^2 + \frac{r_s}{d}\eta_0^3 = \hat{a}_-(d)\,.
\ee

\item
Defining the (signed) distance $\ell$ from the throat as
\be
	\ell(r) = \pm\int_d^r\frac{\td r'}{\sqrt{1 - b_{\pm}(r') / r'}}\,,
\ee
we have the condition that $r(\ell)$ has a minimum at the throat. This in particular implies that
\be
	0 = \left.\frac{\td r}{\td \ell}\right|_{\ell = 0} = \pm\sqrt{1 - \frac{b_{\pm}(d)}{d}}\,,
\ee
and thus $b_{\pm}(d) = d$. Again, it is easy to check that this is the case for the solution we consider. It further implies the so-called flaring-out condition,
\be\label{eqn:throatvic}
	0 < \frac{\td ^2r}{\td \ell^2} = \frac{1}{2r}\left(\frac{b_{\pm}(r)}{r} - b_{\pm}'(r)\right) = \frac{d^2}{r^3}a_{\pm}(r) + \frac{r^2 - d^2}{2r^2}a_{\pm}'(r),
\ee
in the vicinity of the throat. In particular, at the throat itself, where $r = d$ and $b_{\pm}(d) = d$, we must have
\be\label{eqn:throatder}
	0 \leq \left.\frac{d^2r}{d\ell^2}\right|_{\ell = 0} = \frac{1}{2d}\left(1 - b_{\pm}'(d)\right) = \frac{1}{d}a_{\pm}(d)\,.
\ee
However, recall that $a_{\pm}(r) > 0$ in the proximity of the throat as a consequence of the condition that the throat is not surrounded by horizons, so that condition~\eqref{eqn:throatder} is satisfied. From the continuity of $a_{\pm}$ follows that also condition~\eqref{eqn:throatvic} is satisfied.
\end{itemize}

\subsection{Geodesics}
\label{ssec:geodesics}

To clarify whether the wormhole we are considering is traversable, or not, we study geodesics passing through the wormhole. The metric \eqref{stat:g_sss_wh} is spherically symmetric and static so we can use the symmetries to derive the radial timelike and lightlike geodesics in the usual way.

We remark that null geodesics are invariant under conformal transformations. Thus null geodesics of the metric $g$ are identical to null geodesics of the metric $ g$ up to reparametrisation. As the metric \eqref{stat:g_sss_wh}  is conformally related to a region of the Schwarzschild de-Sitter metric, its lightlike geodesics can also be derived from the null geodesics of Schwarzschild de-Sitter spacetime which have been discussed in detail  in~\cite{Hackmann:2008zz}. 

We are looking for curves $\gamma(\tau) = (t(\tau), r(\tau), \theta(\tau), \phi(\tau))$, where $\tau$ denotes the proper time. Without loss of generality we only need to consider curves in the equatorial plane, that is $\theta(\tau)= \pi/2$. The test particles have two constants of motions: energy $E = \hat a \dot t$, where $\hat a$ is defined in \eqref{eqn:geomfunc}, and the angular momentum $\mathcal{L} = 2 (l^2 + d^2)\dot \phi$. The dot denotes derivation with respect to the proper time. The orbit equation is obtained by demanding $g(\dot \gamma, \dot{\gamma}) = \epsilon$, where $\epsilon = -1$ for timelike geodesics and $\epsilon = 0$ for lightlike geodesics. Using the constants of motion we find that the orbit in the wormhole spacetime obeys
\begin{align}\label{eq:geodeq}
	\dot l^2 + 2 V_{\text{eff}}(l) &= E^2\,,
\end{align}
where the effective potential reads
\begin{align}\label{Veff}
	2 V_{\text{eff}}(l) =  \hat a(l)\,\left(\frac{\mathcal{L}}{4 (l^2 + d^2)} - \epsilon \right)\,.
\end{align}

Consider the proper time for a test particle travelling from $l_0 = l(\tau_0)$ to $l_1 = l(\tau_1)$. Eq.~\eqref{eq:geodeq} implies that
\be\label{geo_tau}
	\tau
	=  \int_{l_0}^{l_1} \td l\ \left( E^2 - \left(\frac{\mathcal{L}}{4(l^2 + d^2)} - \epsilon \right)  \hat a \right)^{-1/2}.
\ee
For observers with a sufficiently large energy parameter $E$, namely such that the term under the square root is positive, we obtain a finite proper time even when $l_0$ and $l_1$ lie on different sides of the wormhole throat. If $\hat a$ is positive everywhere, this holds also for the coordinate time
\be\label{geo_t}
	t
	=  \int_{l_0}^{l_1} \ \frac{\td l}{\hat a}\left( 1 - \left(\frac{\mathcal{L}}{4(l^2 + d^2)} - \epsilon \right)  \hat a/E^2 \right)^{-1/2}.
\ee
It is possible that the solution contains turning points defined by a vanishing of the bracket in the integrand. In this case the particle oscillates between the different sides of the wormhole. When the energy exceeds the effective potential in the vincinity of the wormhole throat, it is possible to obtain trajectories passing through the wormhole.

For radially infalling photons ($\mathcal{L}=0$, $\epsilon = 0$) the effective potential \eqref{Veff} vanishes. In terms of the coordinate time $t$ the trajectories then read $\gamma(t) = \left( t, l(t), \pi/2, 0 \right)$ with  $l(t)$ given implicitly by
\be\label{eq:ligtlt}
	t =  \int^{l} \ \frac{\td l'}{\hat{a}(l')} = \int^{d/\eta(l)} \ \frac{\td  r}{a( r)},
\ee
where $a( r)$ is the $g_{00}$ component of the original spherically symmetric spacetime \eqref{eq:kerr-schild}, in our case the $g_{00}$ component of the S-(A)dS spacetime.  It is related to $\hat{a}(l')$ by the wormhole transformation~\eqref{hat_a}, while the coordinates are related by $r = d/\eta(l)$. The identity in~\eqref{eq:ligtlt} follows from the invariance of lightlike geodesics under conformal transformations, and can also be shown by directly applying~\eqref{hat_a}. The possible range of integration is determined by $d/\eta(l)$ and will not be the same for the S-(A)dS spacetime and the wormhole spacetimes.

When inserting the S-(A)dS metric \eqref{stat:g_SdS}, the integral \eqref{eq:ligtlt} can be evaluated analytically. It yields the dependence $l(t)$ which determines the proper time which passes for an observer at rest until a light ray has passed the wormhole.

\subsection{Penrose diagram for the wormhole metric}
\label{ssec:penrose}

Penrose diagrams are a very useful tool to visualise the causal structure of spacetimes. They are constructed by finding a coordinate transformation of the time and radial coordinate $t, l$, such that the new coordinates $T, L$ have a finite range, and that lightlike radial geodesics satisfy $dL/dT = \pm 1$. Such a coordinate transformation can be constructed in several steps, which we list here for the general wormhole metric~\eqref{stat:g_sss_wh}. The results are illustrated in three examples in Section~\ref{ssec:examples}.

We start by introducing the tortoise coordinate $\tilde{l}$ through the definition
\be\label{eq:tortoise}
	\tilde{l}(l) \equiv \int_0^l \frac{\td l'}{\hat{a}(l')}\,,
\ee
while all other coordinates remain unchanged. This is the integral~\eqref{eq:ligtlt} for radial lightlike geodesics and, in general, it has no analytic solution. The coordinate transformation is defined and invertible everywhere, if $a$ is positive, i.e., when the integration variable does not cross any horizons. The metric~\eqref{stat:g_sss_wh} can now be recast
\be
	\td s^2 = \hat{a}(-\td t^2 + \td\tilde{l}^2) + (l^2 + d^2)\td \mathbf{\Sigma}^2\,,
\ee
where $l$ is implicitly defined as a function of $\tilde{l}$. It follows that radial light rays follow curves with $d\tilde{l}/dt = \pm 1$ in these coordinates.

By introducing light cone coordinates $u = t - \tilde{l}$, $v = t + \tilde{l}$ the metric is recast as
\be
	\td s^2 = -\hat{a}\td u\td v + (l^2 + d^2)\td \mathbf{\Sigma}^2\,,
\ee
with $u, v$ being null coordinates. This property is preserved under a coordinate transformation $U = d\tanh\left(u/d\right)$, $V = d\tanh\left(v/d\right)$, after which we obtain null coordinates $U, V$, as can be read off from the metric
\be
	\td s^2 = -\frac{\hat{a}}{(1 - U^2/d^2)(1 - V^2/d^2)} \td U\td V + (l^2 + d^2)\td \mathbf{\Sigma}^2\,.
\ee
Observe that the whole wormhole spacetime is now mapped into the region $U, V \in (-d, d)$ or a subset thereof, depending on the range of the tortoise coordinate $\tilde{l}$, which depends on the choice of the function $\hat{a}$. Hence, also the coordinates
\be
	T = \frac{U + V}{2}\,, \quad L = \frac{V - U}{2},
\ee
take values on a bounded region. The metric then finally takes the form
\be
	\td s^2 = \frac{\hat{a}}{(1 - U^2/d^2)(1 - V^2/d^2)} (-\td T^2 + \td L^2) + (l^2 + d^2)\td \mathbf{\Sigma}^2\,.
\ee
These are the coordinates we used for the Penrose diagrams in section~\ref{ssec:examples}.

\bibliographystyle{JHEP}

\bibliography{CWH_refs}

\end{document}